\documentclass[conference]{IEEEtran}
\IEEEoverridecommandlockouts

\usepackage{cite}
\usepackage{amsmath,amssymb,amsfonts}
\usepackage{algorithmic}
\usepackage{graphicx}
\usepackage{textcomp}
\usepackage{xcolor}

\usepackage{amsbsy}
\usepackage{gensymb}
\usepackage{enumitem}
\usepackage{amsthm}
\usepackage{amsmath,bm}
\usepackage{caption}
\usepackage{subcaption}

\theoremstyle{remark}
\newtheorem{remark}{Remark}

\def\BibTeX{{\rm B\kern-.05em{\sc i\kern-.025em b}\kern-.08em
    T\kern-.1667em\lower.7ex\hbox{E}\kern-.125emX}}
\begin{document}

\title{Transforming Fading Channel from Fast to Slow: IRS-Assisted High-Mobility Communication
}
\author{
\IEEEauthorblockN{Zixuan Huang\textsuperscript{*\dag}, Beixiong Zheng\textsuperscript{*}, and Rui Zhang\textsuperscript{*}}

\IEEEauthorblockA{
\textsuperscript{*}Department of Electrical and Computer Engineering, National University of Singapore, Singapore  117583\\
\textsuperscript{\dag}NUS Graduate School, National University of Singapore, Singapore 119077 \\
Email: huang.zixuan@u.nus.edu; \{elezbe, elezhang\}@nus.edu.sg }

}

\maketitle
\vspace*{-4em}
\begin{abstract}
In this paper, we study a new intelligent refracting surface (IRS)-assisted high-mobility communication with the IRS deployed in a high-speed moving vehicle to assist its passenger's communication with a static base station (BS) on the roadside. The vehicle's high Doppler frequency results in a fast fading channel between the BS and the passenger/user, which renders channel estimation for the IRS with a large number of refracting elements a more challenging task as compared to the conventional case with low-mobility users only. In order to mitigate the Doppler effect and reap the full IRS passive beamforming gain with low training overhead, we propose a new and efficient transmission protocol to execute channel estimation and IRS refraction design for data transmission. Specifically, by exploiting the quasi-static channel between the IRS and user both moving at the same high speed, we first estimate the cascaded BS-IRS-user channel with the Doppler effect compensated. Then, we estimate the instantaneous BS-user fast fading channel (without IRS refraction) and tune the IRS refraction over time accordingly to align the cascaded channel with the BS-user direct channel, thus maximizing the IRS's passive beamforming gain as well as converting their combined channel from fast to slow fading. Simulation results show the effectiveness of the proposed channel estimation scheme and passive beamforming design as compared to various benchmark schemes.
\end{abstract}


\section{Introduction}

Recently, intelligent reflecting/refracting surface (IRS) has emerged as a cost-effective  technique to achieve smart and reconfigurable radio environment for next-generation wireless communication systems \cite{qq1,tutorial}. Specifically, IRS is a digitally-controllable metasurface consisting of a massive number of passive reflecting/refracting elements, whose amplitudes and/or phase shifts can be individually  controlled in real time, thereby enabling  dynamic control over the wireless propagation channel for assorted purposes (e.g., passive beamforming, interference nulling/cancellation \cite{qq1,tutorial}). Moreover, IRS dispenses with radio frequency (RF) chains and only reflects/refracts the ambient signals passively, which features very low hardware cost and energy consumption. As such, IRS has been extensively investigated for various wireless systems and applications such as orthogonal frequency division multiplexing (OFDM) \cite{ofdmi1,ofdmb1,ofdmb2}, non-orthogonal multiple access (NOMA) \cite{normabei,norma1},  and so on.

\begin{figure}[t]
\vspace*{-3em}
\centering
\includegraphics[width=0.43\textwidth]{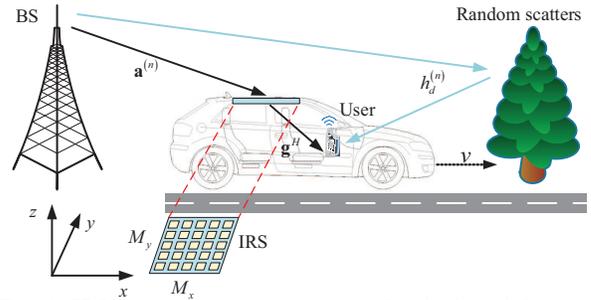}
\vspace*{-0.75em}
\caption{IRS-assisted communication with a high-mobility user.}
\label{config}
\vspace*{-2.2em}
\end{figure}

To fully reap the performance gain of IRS, the acquisition of accurate channel state information (CSI) for the links between the IRS and its associated  base station (BS)/users is crucial, which, however, is practically challenging due to the following reasons. Firstly, the IRS can perform passive signal reflection/refraction only, which renders the CSI acquisition by conventional approaches requiring active transmitter/receiver infeasible \cite{qq1,tutorial}. As such, an alternative approach in practice is to estimate the cascaded BS-IRS-user channels based on the pilot symbols sent by the users/BS with properly designed IRS reflection/refraction pattern over time (see, e.g., \cite{ofdmi1,partichang}). Secondly, IRS generally consists of a vast number of elements, which incur prohibitively high training overhead that may severely reduce the throughput for data transmission. In order to reduce the training overhead, an effective method is to group adjacent reflecting/refracting elements with high spatial channel correlation into a subsurface, whereby only the effective cascaded BS-IRS-user channel associated with each sub-surface (instead of those for its individual elements) needs to be estimated \cite{ofdmi1,partichang}.

However, the existing works on IRS have mostly considered the low-mobility users served by IRS for local coverage only, which is inapplicable to the high-mobility scenario such as high-speed vehicular communication. In this case, the transmitted signal arrives at the receiver over multiple propagation paths with rapidly time-varying phase shifts at different Doppler frequencies in general, which leads to a superimposed fast fading channel (i.e., the received signal's amplitude and phase both vary substantially over time) \cite{gold}. As a result, the achievable data rate and outage probability of the communication system can be greatly degraded. In \cite{bass}, IRS deployed at the roadside was proposed for the high-mobility wireless communication by tuning the IRS reflection to compensate for the severe Doppler effect. Moreover, IRS channel estimation was investigated in \cite{dsce} with the Doppler effect taken into account. However, the above-mentioned studies assumed that the IRS is deployed at a fixed location, which renders very limited time for the IRS to assist the communication between the BS and a high-mobility user passing by it. 

To circumvent this issue, we consider a new IRS-assisted high-mobility communication system in this paper, where the IRS with a large number of refracting elements is deployed in a high-speed moving vehicle to assist the communication between the user (passenger) residing in it and a roadside BS at fixed location, as shown in Fig.~\ref{config}.
To achieve reliable and high-rate communication between the BS and the high-mobility user assisted by the IRS, an efficient transmission protocol is proposed to execute channel estimation and IRS refraction design for data transmission. Specifically, by exploiting the  quasi-static channel between the IRS and user both moving at the same high speed, we first estimate the cascaded BS-IRS-user channel and compensate for the high Doppler frequency between the BS and IRS in the cascaded channel. Then, we estimate the instantaneous BS-user fast fading channel (without IRS refraction) and adjust the IRS refraction accordingly over time to align the cascaded BS-IRS-user channel with the BS-user (direct) channel, thus maximizing the IRS's passive beamforming gain and effectively converting their combined channel from fast to slow fading. It is shown via numerical results that by adopting the proposed efficient channel estimation scheme and IRS refraction design, the IRS-assisted system achieves significant rate improvement over the conventional system with both the direct and cascaded channels estimated over each (short) channel coherence interval in the high-mobility communication scenario.   

{\emph{Notations:}} Upper-case and lower-case boldface letters denote matrices and column vectors, respectively. Upper-case calligraphic letters (e.g., $\mathcal{T}$) denote discrete and finite sets. Superscripts $(\cdot)^{T}$, $(\cdot)^{*}$, $(\cdot)^{H}$, and $(\cdot)^\dagger$ stand for the transpose, conjugate, Hermitian transpose, and Moore-Penrose inverse operations, respectively.
$\left\| \cdot \right\|$ denotes the $l_2$ norm.
$\angle (\cdot)$ denotes the angle of a complex number.
$\mathbb{C}^{x \times y}$ denotes the space of $x \times y$ complex matrices. $\operatorname{diag}(\boldsymbol{x})$ denotes a square diagonal matrix with the elements of $\boldsymbol{x}$ on the main diagonal.
$\otimes$ denotes the Kronecker product. $\odot$ denotes the Hadamard product. 
The distribution of a circularly symmetric complex Gaussian (CSCG) random variable with mean $\mu$ and variance $\sigma^2$ is denoted by $\mathcal{N}_{c}\left(\mu, \sigma^{2}\right)$; and $\sim$ stands for “distributed as”. 
\section{System Model}
As shown in Fig. \ref{config}, we consider an IRS-assisted high-mobility communication system, where an IRS is deployed on the top of a high-speed vehicle to assist its passenger/user's communication with a static BS. 
We assume that the vehicle moves at a high speed of $v$ meters/second (m/s). For the purpose of exposition, we consider the donwlink communication and assume that both of the BS and user terminal are equipped with a single antenna. Moreover, we assume that the IRS is a uniform planar array (UPA) composed of $M_\mathrm{R} = M_x \times M_y$ refracting elements placed in the $x-y$ plane in the three-dimensional (3D) Cartesian coordinate system, which is connected to a smart controller that adjusts its refracting  elements’ individual amplitudes and/or phase shifts by exchanging required information with the BS/user.
We focus on one transmission frame of duration $T$, which is divided into $N$ time blocks, each with an equal duration of $T_c = T/N$.


Let $h^{\left( n \right)}_d$ denote the baseband equivalent channel for the direct link from the BS to the user without any IRS refraction during block $n \in \mathcal{N} \triangleq\left\{1, \ldots, N\right\}$, which is assumed to remain constant during each block, but may change randomly from one block to another.
Let $\mathbf{s}\left( \phi, M \right) = \left[1,e^{-j \pi \phi}, ..., e^{-j( M-1 )\pi \phi}\right]^T$ denote the one-dimensional (1D) steering vector function of the IRS (for both its receiving and refracting) which is applicable to each row/column of its elements, where $\phi$ denotes the phase difference (normalized to $\pi$) between two adjacent elements in the row/column and $M$ denotes the number of elements involved.
Denote the BS-IRS channel and IRS-user channel by $\mathbf{a}^{\left( n \right)} \in \mathbb{C}^{M_\mathrm{R} \times 1}$ and $\mathbf{g}^H \in \mathbb{C}^{M_\mathrm{R} \times 1}$, respectively, which are both assumed to be line-of-sight (LoS) for simplicity\footnote{The results in this paper can be extended to the general case with multiple signal paths (LoS and non-LoS) between the IRS and BS/user, which will be considered in our future work.}. Due to the high-speed moving vehicle, the BS-IRS link is subjected to the Doppler frequency and its equivalent baseband channel is given by 
\begin{equation}\label{bs-irs}
\vspace*{-1.2em}
\mathbf{a}^{\left( n \right)} = \alpha_\mathrm{BI} e^{j2\pi f_d n T_c} \underbrace{\mathbf{s}\left( \phi_\mathrm{BI}, M_x \right) \otimes \mathbf{s}\left( \varphi_\mathrm{BI}, M_y \right)}_{\mathbf{u}_\mathrm{BI}\left(\theta_\mathrm{BI} , \vartheta_\mathrm{BI}\right)},
\vspace*{0.4em}
\end{equation} 
where $\alpha_\mathrm{BI}$ denotes the complex-valued path gain of the BS-IRS link, $f_d = v \cos \theta_\mathrm{BI} \cos \vartheta_\mathrm{BI} /\lambda$ is the Doppler frequency with $\lambda$ being the signal wavelength, 
$\theta_\mathrm{BI} \in \left[0,\pi/2\right]$ and $\vartheta_\mathrm{BI} \in \left[0,2\pi\right)$ denote the elevation and azimuth angles-of-arrival (AoAs) at the IRS, respectively, and $\mathbf{u}_\mathrm{BI}\left(\theta_\mathrm{BI} , \vartheta_\mathrm{BI}\right) \in \mathbb{C}^{M_\mathrm{R} \times 1}$ represents the receive array response vector of IRS, where $\phi_\mathrm{BI} \triangleq \frac{2 d}{\lambda} \cos \theta_\mathrm{BI} \cos \vartheta_\mathrm{BI} \in \left[-\frac{2 d}{\lambda}, \frac{2 d}{\lambda}\right]$ and $\varphi_\mathrm{BI} \triangleq \frac{2 d}{\lambda} \cos \theta_\mathrm{BI} \sin \vartheta_\mathrm{BI} \in \left[-\frac{2 d}{\lambda}, \frac{2 d}{\lambda}\right]$ with $d$ denoting the horizontal/vertical spacing between two adjacent IRS elements.
For the purpose of exposition, we assume that the channel parameters $\alpha_\mathrm{BI}$, $f_d$, $\theta_\mathrm{BI}$, and $\vartheta_\mathrm{BI}$ remain approximately constant for the BS-IRS link in (\ref{bs-irs}) within one transmission frame of interest. 
This assumption is practically valid if the vehicle speed $v$ is constant during one transmission frame and its traveling distance within one transmission frame is negligible as compared to the nominal distance with the BS \cite{gao1}.
On the other hand, the IRS-user link is modeled as an LoS channel given by
\begin{equation}\label{irs-user}
\vspace{-1.5em}
\mathbf{g} = \alpha_\mathrm{IU} 
\underbrace{\mathbf{s}\left( \phi_\mathrm{IU}, M_x \right) \otimes \mathbf{s}\left( \varphi_\mathrm{IU}, M_y \right)}_{\mathbf{u}_\mathrm{IU}\left(\theta_\mathrm{IU}, \vartheta_\mathrm{IU}\right)},
\vspace{0.8em}
\end{equation} 
where $\alpha_\mathrm{IU}$ denotes the complex-valued path gain of the IRS-user link,
$\theta_\mathrm{IU} \in \left[-\pi/2,0\right]$ and $\vartheta_\mathrm{IU} \in \left[0,2\pi\right)$
denote the elevation and azimuth angles-of-departure (AoDs) from the IRS to the user, respectively, and $\mathbf{u}_\mathrm{IU}\left(\theta_\mathrm{IU}, \vartheta_\mathrm{IU}\right)$ represents the refraction array response vector of IRS with $\phi_\mathrm{IU} \triangleq \frac{2 d}{\lambda} \cos \theta_\mathrm{IU} \cos \vartheta_\mathrm{IU} \in \left[-\frac{2 d}{\lambda}, \frac{2 d}{\lambda}\right]$ and $\varphi_\mathrm{IU} \triangleq \frac{2 d}{\lambda} \cos \theta_\mathrm{IU} \sin \vartheta_\mathrm{IU} \in \left[-\frac{2 d}{\lambda}, \frac{2 d}{\lambda}\right]$. 
Note that the IRS-user channel $\mathbf{g}$ remains constant within the considered transmission frame based on the assumption that the IRS and user remain relatively static\footnote{Without loss of generality, we can also set the frame duration $T$ as the minimum time during which $\alpha_\mathrm{BI}$, $f_d$, $\theta_\mathrm{BI}$, and $\vartheta_\mathrm{BI}$ as well as $\mathbf{g}$ remain unchanged based on the practical setup.}.
Let $\boldsymbol{\Omega}^{\left( n \right)} = \operatorname{diag}\left(e^{j \omega^{\left( n \right)}_{1}},  \ldots,e^{j \omega^{\left( n \right)}_{M}}\right) \in \mathbb{C}^{M_\mathrm{R} \times M_\mathrm{R} }$ denote the diagonal IRS refraction matrix at block $n$, where the refraction amplitude of each element is set to one (unless otherwise stated) to maximize the signal refraction power, and $\omega^{\left( n \right)}_{m}$, $m \in\left\{1,2, \ldots, M\right\}$, denotes the refraction phase shift of element $m$. 
The BS-IRS-user channel with the IRS refraction taken into account, denoted by $h^{\left( n \right)}_r $, is thus expressed as
\begin{align}\label{hr}
h^{\left( n \right)}_r &= \mathbf{g}^{H} \boldsymbol{\Omega}^{\left( n \right)} \mathbf{a}^{\left( n \right)}  \nonumber\\
&=\alpha_\mathrm{BI} \alpha_\mathrm{IU} e^{j2\pi f_d n T_c} \mathbf{u}^H_\mathrm{IU}\left(\theta_\mathrm{IU}, \vartheta_\mathrm{IU}\right) \boldsymbol{\Omega}^{\left( n \right)} \mathbf{u}_\mathrm{BI}\left(\theta_\mathrm{BI}, \vartheta_\mathrm{BI}\right) \nonumber\\
&= \left(\mathbf{c}^{\left( n \right)}\right)^H  \mathbf{v}^{\left( n \right)}, \quad \forall n \in \mathcal{N},  
\end{align}
where $\mathbf{v}^{\left( n \right)} = \left[e^{j \omega^{\left( n \right)}_{1}},  \ldots,e^{j \omega^{\left( n \right)}_{M}}\right] ^T$ denotes the IRS refraction vector and $\left(\mathbf{c}^{\left( n \right)}\right)^H$ denotes the cascaded BS-IRS-user channel (without IRS phase shifts) at block $n$, which is expressed as 
\begin{align}\label{stee}
\left(\mathbf{c}^{\left( n \right)}\right)^H &= \beta e^{j2\pi f_d n T_c}\mathbf{u}^H_\mathrm{IU}\left(\theta_\mathrm{IU}, \vartheta_\mathrm{IU}\right) \odot \mathbf{u}^T_\mathrm{BI}\left(\theta_\mathrm{BI}, \vartheta_\mathrm{BI}\right) \nonumber\\
&=\beta e^{j2\pi f_d n T_c} \left(\mathbf{s}^H\left( \phi_\mathrm{IU}, M_x \right) \otimes \mathbf{s}^H\left( \varphi_\mathrm{IU}, M_y \right)\right) \nonumber\\
& \hspace{1.4em} \odot \left(\mathbf{s}^T\left( \phi_\mathrm{BI}, M_x \right) \otimes \mathbf{s}^T\left( \varphi_\mathrm{BI}, M_y \right)\right) \nonumber\\
&\stackrel{(a)}{=}\beta e^{j2\pi f_d n T_c}\left(\mathbf{s}^H\left( \phi_\mathrm{IU}, M_x \right) \odot \mathbf{s}^T\left( \phi_\mathrm{BI}, M_x \right)  \right) \nonumber\\
&  \hspace{1.4em}  \otimes \left(\mathbf{s}^H\left( \varphi_\mathrm{IU}, M_y \right) \odot \mathbf{s}^T\left( \varphi_\mathrm{BI}, M_y \right)  \right)  \nonumber\\
&= \beta e^{j2\pi f_d n T_c} \mathbf{s}^H \left( \tilde{\psi}_\mathrm{x}, M_x \right) \otimes \mathbf{s}^H \left( \tilde{\psi}_\mathrm{y}, M_y \right),
\end{align}
where $(a)$ is obtained according to the mixed-product property of Kronecker product, $\beta \triangleq \alpha_\mathrm{BI} \alpha_\mathrm{IU}$ denotes the product path gain of the cascaded BS-IRS-user link, $\tilde{\psi}_\mathrm{x}=\phi_\mathrm{IU}-\phi_\mathrm{BI} \in \left[-\frac{4 d}{\lambda}, \frac{4 d}{\lambda}\right]$, and $\tilde{\psi}_\mathrm{y}=\varphi_\mathrm{IU}-\varphi_\mathrm{BI} \in \left[-\frac{4 d}{\lambda}, \frac{4 d}{\lambda}\right]$. Then, by noting that $\mathbf{s}\left( \phi, M \right)$ is a periodic function of $\phi$ with period 2, we define $\psi_\mathrm{x} \triangleq \tilde{\psi}_\mathrm{x} \left(\bmod \ 2\right) \in \left[-1, 1\right]$ $\left(\psi_\mathrm{y} \triangleq \tilde{\psi}_\mathrm{y} \left(\bmod \ 2\right) \in \left[-1, 1\right]\right)$ as the cascaded BS-IRS-user channel effective phase along the $x$-axis ($y$-axis), such that we have $\mathbf{s} \left(\psi_\mathrm{x}, M_x \right) = \mathbf{s} \left( \tilde{\psi}_\mathrm{x}, M_x \right)$ and $\mathbf{s} \left( \psi_\mathrm{y}, M_y \right)= \mathbf{s} \left( \tilde{\psi}_\mathrm{y}, M_y \right)$. For the purpose of exposition, we assume $M_y = 1$, while the results of the this paper can be readily extended to the case with $M_y > 1$.
As a result, $\mathbf{c}^{\left( n \right)}$ in (\ref{stee}) can be simplified as
\begin{align}\label{cn}
\mathbf{c}^{\left( n \right)} = \beta e^{j2\pi f_d n T_c} \mathbf{s} \left(\psi_\mathrm{x}, M_x \right),
\end{align}
and in the sequel we mainly focus on the transmission design over the cascaded BS-IRS-user channel along the $x$-axis with $\psi_\mathrm{x}$ considered in (\ref{stee}) only. For notation convenience, we omit $M_x$ in (\ref{cn}) in the rest of this paper.
Hence, the effective channel between the BS and user by combining the BS-user direct channel and the BS-IRS-user cascaded channel is given by 
\begin{align}
\tilde{h}^{\left( n \right)} 
&= \left(\mathbf{c}^{\left( n \right)}\right)^H  \mathbf{v}^{\left( n \right)} +h^{\left( n \right)}_d 
\label{equiva:1}
\\
&= \beta e^{j2\pi f_d n T_c} \mathbf{s}^H \left(\psi_\mathrm{x}\right) \mathbf{v}^{\left( n \right)} +h^{\left( n \right)}_d.
\label{equiva:2}
\end{align}
Note that in (\ref{equiva:2}), the cascaded BS-IRS-user channel $\mathbf{c}^{\left( n \right)}$ varies over different blocks due to the Doppler-induced phase shift $e^{j2\pi f_d n T_c}$.
In contrast, the BS-user direct channel $h^{\left( n \right)}_d$, which is the superimposition of all the non-IRS-refracted paths, varies over different blocks in both amplitude and phase (i.e., fast fading). 
Thus, the different time-varying behaviors of $\mathbf{c}^{\left( n \right)}$ and $h^{\left( n \right)}_d$ motivate us to first estimate the BS-IRS-user cascaded channel as well as the Doppler frequency $f_d$ so as to compensate for the Doppler-induced phase variation in it, and then estimate the fast fading channel $h^{\left( n \right)}_d$ at each time block $n$ and adjust the IRS refraction to achieve passive beamforming toward the user as well as coherent combining of the signals over the direct and cascaded channels to maximize the received signal power at the user. This not only leads to an efficient transmission protocol for channel estimation and IRS refraction design for data transmission, but also effectively converts the high-mobility induced fast fading channel (without IRS) to one with much more slowly time-varying amplitude and phase (i.e., slow fading), by optimally tuning the IRS refraction over time.     
Note that the optimal design of IRS refraction vector $\mathbf{v}^{\left( n \right)}$ at each time block $n$ for data transmission requires the CSI of the Doppler frequency $f_d$, the cascaded BS-IRS-user channel phase $\psi_\mathrm{x}$ and path gain $\beta$, as well as the instantaneous BS-user direct channel $h^{\left( n \right)}_d$ of each block.
To this end, we propose a new transmission protocol to perform channel estimation, Doppler compensation, and IRS refraction design for enhancing data transmission, as elaborated in the next section.

\section{Proposed Transmission Protocol}
\begin{figure}[h]
\vspace*{-1.1em}
\centering
\includegraphics[width=0.42\textwidth]{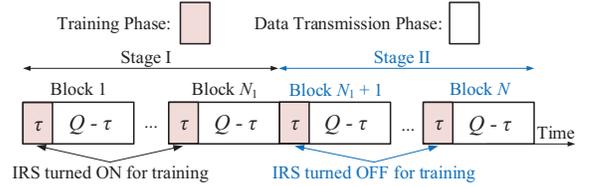}
\vspace*{-0.2em}
\caption{Illustration of the proposed transmission protocol.}
\label{protocol}
\vspace*{-0.9em}
\end{figure}
In this section, we present a new two-stage transmission protocol for the IRS-assisted high-mobility communication, as illustrated in Fig. \ref{protocol}. Specifically, each transmission frame of $N$ blocks is divided into two stages, referred to as Stage~I and Stage~II, comprising $N_1$ and $N-N_1$ blocks, respectively. 
Moreover, each block comprises $Q$ symbols with equal duration of $T_s = T_c /Q$, which are further divided into $\tau$ pilot symbols and $Q-\tau$ data symbols. 
In the following, we elaborate the two transmission stages and their corresponding training and data transmission phases per block, respectively.
\subsection{Stage~I}
\subsubsection{Training Phase}
Let $ \mathcal{N}_\mathrm{I} \triangleq\{1 , \ldots, N_1\}$ denote the index set of the first $N_1$ blocks in Stage~I. 
First, for the training phase of each block $n \in \mathcal{N}_\mathrm{I}$, we dynamically tune the refraction of the IRS, $\mathbf{v}^{\left(n\right)}_i$, over different pilot symbols to facilitate the channel estimation, where $i\in \mathcal{T} \triangleq\{1 , \ldots, \tau\}$. 
Based on the channel model in (\ref{equiva:1}), the received signal at the user during pilot symbol $i$ of block $n$ can be expressed as
\begin{align}\label{received_sym1}
y^{\left(n\right)}_i = \left(\left(\mathbf{c}^{\left( n \right)}\right)^H \mathbf{v}^{\left(n\right)}_i + h^{\left( n \right)}_d\right) x^{\left( n \right)}_i + z^{\left( n \right)}_i, n\in \mathcal{N}_\mathrm{I}, i\in \mathcal{T},
\vspace*{-1em}
\end{align}
where $x^{\left( n \right)}_i$ denotes the pilot symbol transmitted by the BS which is simply set as $x^{\left( n \right)}_i = 1$ for ease of exposition,  $z^{\left( n \right)}_i \sim\mathcal{N}_{c}\left(0, \sigma^{2}\right)$ is the received noise with $\sigma^{2}$ being the normalized noise power (with respect to transmit power). 
For simplicity, we design the IRS training refraction vector as $\mathbf{v}^{\left(n\right)}_i =  \mu_i \bar{\mathbf{v}}^{\left(n\right)}$ during pilot symbol $i$, where $\bar{\mathbf{v}}^{\left(n\right)}$ denotes the initial training refraction vector in block $n$ and $\mu_i \in \mathbb{C}$ with $\left|\mu_i\right| = 1$ denotes the time-varying common phase shift that is applied to all the refracting elements of the IRS during the training phase of each block.
In this paper, we consider the random phase-shift design for $\bar{\mathbf{v}}^{\left(n\right)}$, where the training refraction phase-shifts of $\bar{\mathbf{v}}^{\left(n\right)}$ are randomly generated following the uniform distribution within $\left[0,2 \pi\right)$.
As such, the resultant received signal in (\ref{received_sym1}) can be simplified as
\begin{align}\label{received_sym11}
\vspace*{-0.7em}
y^{\left(n\right)}_i 
= \mu_i \bar{h}^{\left( n \right)}_r + h^{\left( n \right)}_d  + z^{\left( n \right)}_i , \quad n\in \mathcal{N}_\mathrm{I},i\in \mathcal{T},
\vspace*{-0.7em}
\end{align}
where $\bar{h}^{\left( n \right)}_r  =  \left(\mathbf{c}^{\left(n\right)}\right)^H \bar{\mathbf{v}}^{\left(n\right)}$ denotes the initial IRS-refracted channel of each block.
By collecting $\tau$ received pilot symbols $\{y^{\left(n\right)}_i\}_{i=1}^\tau$ for each block, the received signal vector can be expressed as 
\begin{equation}\label{received_ce}
\mathbf{y}^{\left(n\right)} = \mathbf{\Theta} \mathbf{h}^{\left(n\right)} + \mathbf{z}^{\left(n\right)}, \quad n\in \mathcal{N}_\mathrm{I},
\end{equation}
where  $\mathbf{\Theta} = \left[ \bar{\bm{\mu}}_1, \ldots,  \bar{\bm{\mu}}_\tau  \right]^T \in \mathbb{C}^{\tau \times 2}$ denotes the training  matrix for all blocks
with $\bar{\bm{\mu}}_i = \left[1, \mu_i\right]^T $, $\mathbf{h}^{\left(n\right)} = \left[h^{\left( n \right)}_d , \bar{h}^{\left( n \right)}_r  \right]^T$ denotes the channel vector including both the IRS-refracted and non-IRS-refracted channels, and $\mathbf{z}^{\left(n\right)} = \left[z^{\left(n\right)}_{1}, \ldots, z^{\left(n\right)}_{\tau}\right]^{T}$ denotes the  received noise vector. 
By properly constructing the training  matrix $\mathbf{\Theta}$ such that $\mathrm{rank} \left(\mathbf{\Theta}\right) = 2$, the least-square (LS) estimate of $\mathbf{h}^{\left(n\right)}$ based on (\ref{received_ce}) is given by
\begin{equation}\label{est_bl}
\hat{\mathbf{h}}^{\left(n\right)} 
=\mathbf{\Theta}^\dagger
\mathbf{y}^{\left(n\right)} = \mathbf{h}^{\left(n\right)} + \mathbf{\Theta}^\dagger \mathbf{z}^{\left(n\right)}, \quad n\in \mathcal{N}_\mathrm{I}.
\end{equation}
Note that $\tau \ge 2$ is required to ensure $\mathrm{rank} \left(\mathbf{\Theta}\right) = 2$ and thus the existence of $ {\bm \Theta}^{\dagger}$. 
Similar to \cite{ofdmb1,ofdmb2}, we can set the training matrix $\mathbf{\Theta}$ as the submatrix of the $\tau\times  \tau$ discrete Fourier transform (DFT) matrix with its first two columns.

\subsubsection{Data Transmission Phase}
With the estimated CSI $\hat{\mathbf{h}}^{\left(n\right)}$ and the refraction vector set as the one among $\{\mathbf{v}^{\left( n \right)}_i\}_{i=1}^\tau$, denoted by $\mathbf{v}^{\left( n \right)}_\mathrm{I,D}$, which achieves the maximum received signal power, the user decodes the data in the remaining $Q-\tau$ symbols of each block. Thus, the achievable rate (with the training overhead taken into account in each block) in bits per second per Hertz (bps/Hz) of Stage~I is given by 
\begin{align}\label{rate1}
\vspace*{-1em}
R_\mathrm{I} = \frac{Q -\tau}{N_1 Q} \sum_{n=1}^{N_1} \log _{2}\left(1+\frac{ W_\mathrm{I}^{\left( n \right)} }{\Gamma \sigma^{2}}\right), 
\vspace*{-0.5em}
\end{align}
where $W_\mathrm{I}^{\left( n \right)}  = \left|\left(\mathbf{c}^{\left( n \right)}\right)^{H} \mathbf{v}^{\left( n \right)}_\mathrm{I,D}+h_{d}^{\left( n \right)} \right|^2$ denotes the channel gain of block $n$ in Stage~I for data transmission and $\Gamma \ge 1$ denotes the achievable rate gap due to a practical modulation and coding scheme.
At the end of Stage~I, the user collects the estimated CSI $\{\hat{\bar{h}}^{\left( n \right)}_r\}^{N_1}_{n=1}$ in (\ref{est_bl}) over $N_1$ blocks, and then jointly estimates the channel parameters associated with the BS-IRS-user link (i.e., $\{f_d, \psi_\mathrm{x}, \beta\}$) via the maximum likelihood estimation. For more details on the estimation of $\{f_d, \psi_\mathrm{x}, \beta\}$, please refer to the Appendix. The estimates of $\{f_d, \psi_\mathrm{x}\}$ and that of the phase-shift of $ \beta$, i.e., $\angle \beta$, are then fed back to the IRS controller via a separate control link, which will be used in Stage~II to enhance data transmission.
\subsection{Stage~II}
\subsubsection{Training Phase}
Let $\mathcal{N}_\mathrm{II} \triangleq\{N_1 +1 , \ldots, N\}$ denote the index set of the $N-N_1$ blocks in Stage~II. For the training phase of each block  $n \in \mathcal{N}_\mathrm{II}$, we estimate the BS-user direct channel $h^{\left( n \right)}_d$ in (\ref{equiva:1}) with the IRS turned OFF (i.e.,  $\mathbf{v}^{\left(n\right)}_i= \mathbf{0}$, $\forall i\in \mathcal{T}$) over the $\tau$ pilot symbols.
With $x^{\left( n \right)}_i = 1$ being the pilot symbol transmitted by the BS at pilot symbol $i$ of block $n$, the received signal can be expressed as

\vspace*{-1em}
\begin{align}\label{received_sym2}
y^{\left(n\right)}_i = h^{\left( n \right)}_d + z^{\left( n \right)}_i, \quad  n\in \mathcal{N}_\mathrm{II},i\in \mathcal{T},
\end{align}
where $z^{\left( n \right)}_i \sim\mathcal{N}_{c}\left(0, \sigma^{2}\right)$ is the normalized   received noise. 
Then, the estimate of $h^{\left( n \right)}_d$ is obtained as 

\vspace*{-1em}
\begin{align}\label{esti_hd}
\vspace*{-2em}
\hat{h}^{\left( n \right)}_d = \frac{1}{\tau} \sum_{i=1}^{\tau} y^{\left(n\right)}_i , \quad n\in \mathcal{N}_\mathrm{II}.
\vspace*{-2em}
\end{align}
\vspace*{-0.75em}

After that, the phase-shift of $\hat{h}^{\left( n \right)}_d$, i.e., $\angle \hat{h}^{\left( n \right)}_d$, is fed back to the IRS controller via the control link at the end of the training phase of each block.\footnote{For simplicity, we assume that the control link is reliable with negligible delay; while the proposed protocol can be modified to accommodate small delay in practice by exploiting the channel correlation between consecutive blocks in practice.} 
\subsubsection{Data Transmission Phase}
Denote the IRS refraction vector for data transmission at block $n$ as $\mathbf{v}^{\left( n \right)}_\mathrm{II,D}$.
Based on the estimated CSI of $\hat{h}^{\left( n \right)}_d$ in (\ref{esti_hd}) as well as those of $\{  \hat{f}_d, \hat{\psi}_\mathrm{x} , \hat{\beta}\}$ acquired at the end of Stage~I, the estimated effective channel for data transmission in block $n$ is given by
\begin{equation}\label{w2}
\vspace*{-0.5em}
\hat{\tilde{h}}^{\left( n \right)}_\mathrm{II,D}  =
\hat{\beta} e^{j2\pi \hat{f}_d n T_c} \mathbf{s}^H \left(\hat{\psi}_\mathrm{x}\right) \mathbf{v}^{\left( n \right)}_\mathrm{II,D} + \hat{h}^{\left( n \right)}_d , \quad n\in \mathcal{N}_\mathrm{II}.
\end{equation}
Note that the channel power of $\hat{\tilde{h}}^{\left( n \right)}_\mathrm{II,D}$ in (\ref{w2}) can be maximized by properly designing $\mathbf{v}^{\left( n \right)}_\mathrm{II,D}$ to compensate for the Doppler-induced phase shift and maximize the passive beamforming gain. 
To this end, we design the IRS refraction vector $\mathbf{v}^{\left( n \right)}_\mathrm{II,D}$ for data transmission in each block as
\begin{equation}\label{v_design}
\mathbf{v}^{\left( n \right)}_\mathrm{II,D}
= e^{-j \left( 2\pi \hat{f}_d n T_c + \angle \hat{\beta} - \angle \hat{h}^{\left( n \right)}_d \right) }  \mathbf{s} \left(\hat{\psi}_\mathrm{x}\right), \quad n\in \mathcal{N}_\mathrm{II}.
\end{equation}
Note that only $\angle \hat{h}^{\left( n \right)}_d$ needs to be fed back from the user to IRS for implementing $\mathbf{v}^{\left( n \right)}_\mathrm{II,D}$ in each block $n \in \mathcal{N}_\mathrm{II}$.   
With the IRS refraction vector $\mathbf{v}^{\left( n \right)}_\mathrm{II,D}$ set as in (\ref{v_design}) and the estimated effective channel in (\ref{w2}), the user decodes the data in the remaining $Q-\tau$ symbols of each block.
Accordingly, the achievable rate of Stage~II is given by 
\begin{align}\label{rate2}
\vspace*{-0.8em}
R_\mathrm{II} = \frac{Q -\tau}{\left(N - N_1\right) Q} \sum_{n= N_1 + 1}^{N} \log _{2}\left(1+\frac{ W_\mathrm{II}^{\left( n \right)}}{\Gamma \sigma^{2}}\right),
\vspace*{-0.5em}
\end{align}
where $W_\mathrm{II}^{\left( n \right)} = \left|\left(\mathbf{c}^{\left( n \right)}\right)^{H} \mathbf{v}^{\left( n \right)}_\mathrm{II,D}+h_{d}^{\left( n \right)} \right|^2$ denotes the channel gain of block $n$ in Stage~II for data transmission. 
Note that $W_\mathrm{II}^{\left( n \right)}$ in Stage II is significantly larger than $W_\mathrm{I}^{\left( n \right)}$ in Stage I on average due to the IRS passive beamforming gain and coherent signal combining over the cascaded and direct links.  
With the achievable rates given in (\ref{rate1}) and (\ref{rate2}) for Stages~I~and~II, respectively, the overall achievable rate of the transmission frame is obtained as
\begin{align}\label{rateover}
\vspace*{-1em}
R =  \frac{N_1}{N}  R_\mathrm{I} + \frac{N -N_1}{N}  R_\mathrm{II}.
\vspace*{-1em}
\end{align}

\begin{remark}
For the proposed transmission protocol with a fixed number of blocks in each  transmission  frame,  an intuitive impact of the number of blocks assigned to Stage~I, i.e., $N_1$, can be envisioned as follows: with too small $N_1$, the estimates of $\{f_d, \psi_\mathrm{x}, \angle \beta\}$ are not accurate enough for the IRS refraction design and thus the passive beamforming gain is degraded in Stage~II, while too large $N_1$ results in less time for reaping the high passive beamforming gain for data transmission in Stage~II; both will reduce the overall  achievable rate.
Therefore, there exists a fundamental trade-off in the time/block allocation between the two stages for maximizing the achievable rate, as will be shown by simulations in Section IV.
\end{remark}

\vspace*{-0.75em}
\section{Simulation Results}
In this section, we examine the performance of our proposed transmission protocol via numerical results.
We set the carrier frequency as $f_c = 5.9$~GHz, and the signal bandwidth as 500~KHz. The vehicle speed is set as $v=50$~m/s, which results in a Doppler frequency with the maximum value of $f_{max} = v f_c/c  \approx 1 $~KHz with $c$ being the speed of light.
Each frame consists of $N = 30$ blocks and the duration of each block is set as $T_c = 1/\left(5 f_{max}\right)  \approx  0.2$~ms, during which the BS-user direct channel is assumed to remain approximately constant.
We set half-wavelength spacing for the adjacent IRS refracting elements.
The path loss exponents of the BS-user, BS-IRS, IRS-user links are set as 3, 2.3, and 2.2, respectively, and the  channel power gain at the reference distance of 1~m is set as $\xi_0 = -30$~dB for each individual link. The distance between the BS and IRS/user is set as 500~m and that between the IRS and user is set as 1.5~m.
The elevation and azimuth AoAs at the IRS are set as $\theta_\mathrm{BI} = 60^{\circ}$ and $\vartheta_\mathrm{BI} = 0^{\circ}$, respectively; while the elevation and azimuth AoDs of the IRS-user link are set as
$\theta_\mathrm{IU} = -45^{\circ}$ and $\vartheta_\mathrm{IU} = 0^{\circ}$, respectively.
We assume that the BS-user direct channel $h_{d}^{\left( n \right)}$ for $n\in \mathcal{N}$ follows the Rayleigh fading due to the blockage of the IRS with its time correlation following the Jake's spectrum \cite{gold}.
Let $P_t$ denote the transmit power at the BS and the noise power at the user is set as $\sigma_u^2 = -110$ dBm. Accordingly, the normalized noise power at the user is given by $\sigma^2 = \sigma_u^2/P_t$.
For each block, we set $\tau = 2$ and the channel capacity gap is set as  $\Gamma = 9$ dB.
For our proposed transmission protocol, we define $\gamma^{\left(n\right)} = W^{\left( n \right)}/\sigma^2$ as the effective channel signal-to-noise ratio (SNR) for data transmission at block $n\in \mathcal{N}$, with $W^{\left( n \right)} = W_\mathrm{I}^{\left( n \right)}$ for $n \in \mathcal{N}_\mathrm{I}$ and $W^{\left( n \right)} = W_\mathrm{II}^{\left( n \right)}$ for $n \in \mathcal{N}_\mathrm{II}$. We first consider the IRS random refraction (RR) design as the benchmark scheme, where we set $\tau = 2$ and the phase-shifts of the initial training refraction vector $\bar{\mathbf{v}}^{\left(n\right)}$ in each block $n\in \mathcal{N}$ are randomly generated following the uniform distribution within $[0,2\pi)$ for estimating the effective channels over $\tau$ pilot symbols. Then the IRS draws the best refraction vector among $\{\mathbf{v}^{\left( n \right)}_i\}_{i=1}^\tau$ that achieves the maximum effective channel gain for data transmission in block $n$. Note that the RR scheme can be considered as a special case of our proposed scheme with $N_1 = N$. 

\begin{figure}
     \begin{subfigure}[h]{0.24\textwidth}
         \centering
         \includegraphics[width=\textwidth]{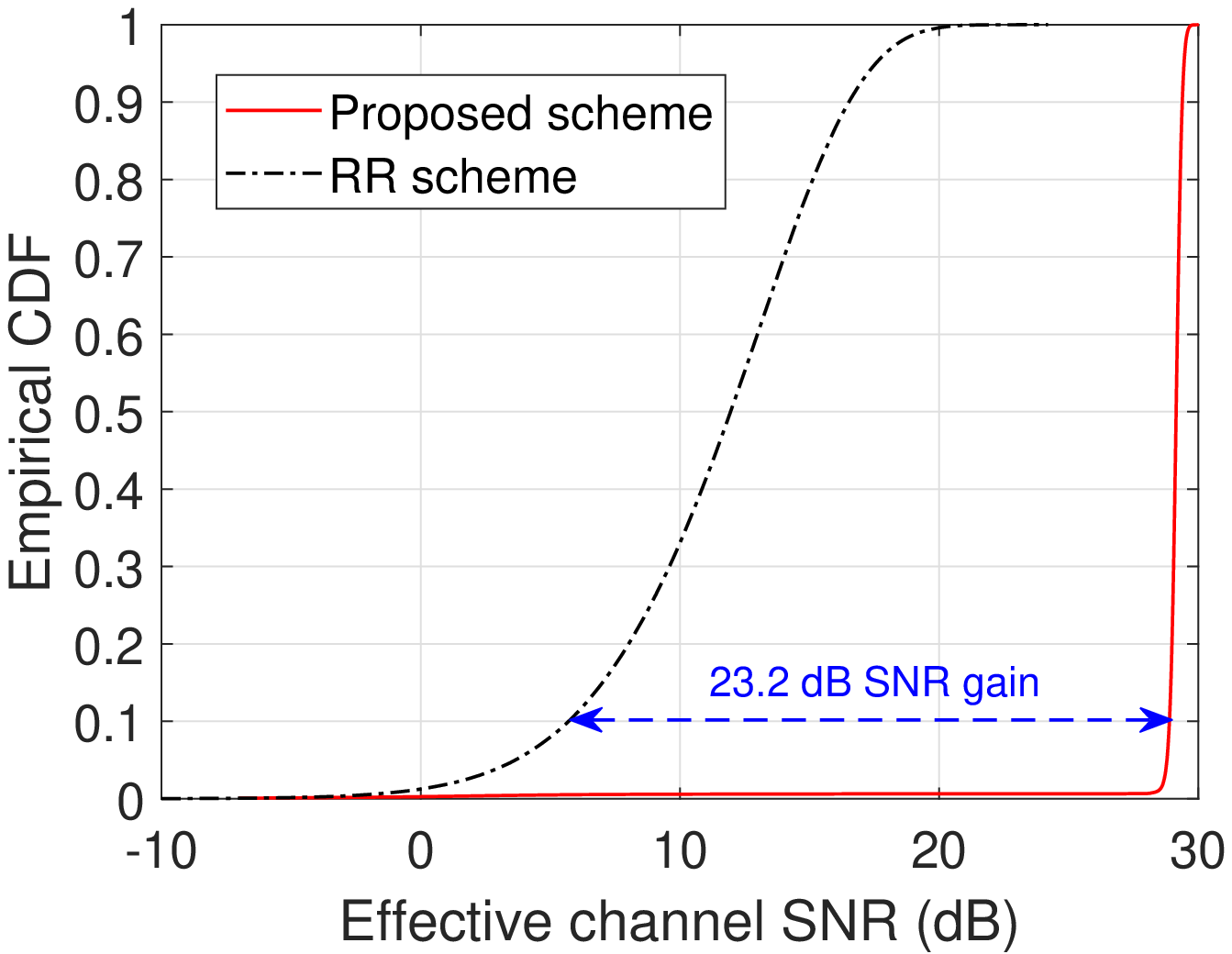}
         \vspace*{-1.55em}
         \caption{Empirical CDF of the effective channel SNR.}
     \end{subfigure}
     \hfill
     \begin{subfigure}[h]{0.24\textwidth}
         \centering
         \includegraphics[width=\textwidth]{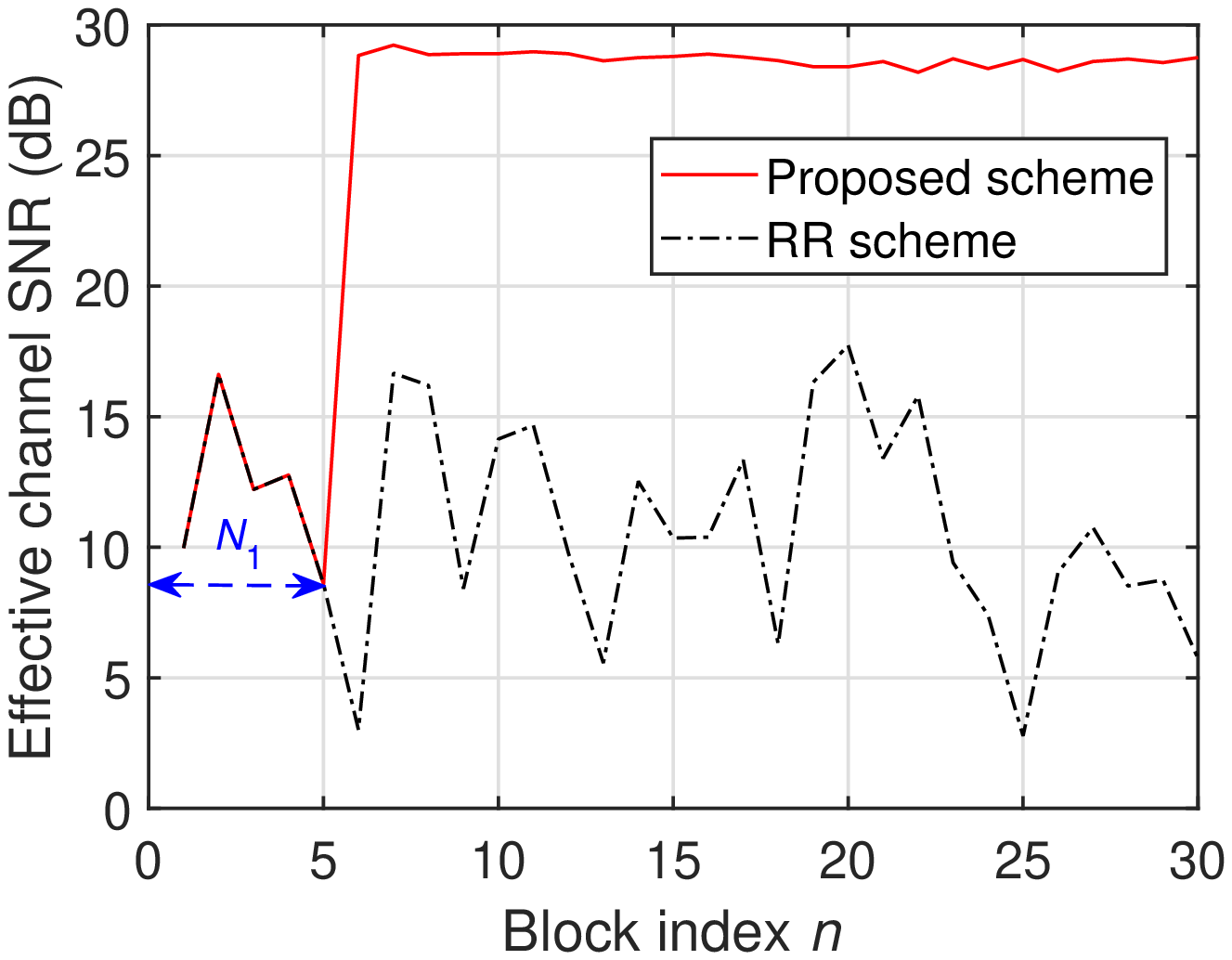}
         \vspace*{-1.55em}
         \caption{A realization of the effective channel SNR in a transmission frame.}
     \end{subfigure}
     \vspace*{-0.5em}
     \caption{Performance comparison of the proposed and RR schemes.}
     \vspace*{-0.5em}
\end{figure}

In Fig.~3, we demonstrate the effectiveness of the proposed scheme in converting the effective channel from fast to slow fading, with $P_t = 41$ dBm, $N_1 = 5$, and $M_\mathrm{R} = 50$. 
In Fig.~3(a), we show the empirical cumulative distribution function (CDF) of the effective channel SNR, $\gamma^{\left(n\right)}$, of the proposed  and  RR schemes, for $n \in \mathcal{N}_\mathrm{II}$. It is observed that due to the IRS passive beamforming gain in Stage~II, the proposed scheme achieves up to 23.2~dB SNR gain over the RR scheme at the same outage rate of 10\%. In Fig.~3(b), we show one realization of $\gamma^{\left(n\right)}$, for $n\in \mathcal{N}$. It is observed that when $n>5$ (i.e., $n \in \mathcal{N}_\mathrm{II}$), the proposed scheme achieves not only much higher average SNR but also significantly less channel gain fluctuation (i.e, slower fading) in Stage~II, as compared to the RR scheme as well as in Stage I. 

Next, we consider the achievable rate comparison with the following additional benchmark schemes: 
\begin{enumerate}
  \item \textbf{Proposed Scheme without (w/o) Channel Phase Alignment (CPA)}: In this scheme, we only perform Doppler compensation and passive beamforming for the IRS-refracted link but without coherently aligning it with the non-IRS-refracted link in Stage~II (i.e., by omitting the term $\angle \hat{h}^{\left( n \right)}_d$ in (\ref{v_design})). 
  \item \textbf{Conventional Cascaded Channel Estimation (CCCE)}: The cascaded channels associated with all refracting elements and the non-IRS-refracted channel are estimated for each block as in \cite{ofdmb1}, based on which the IRS performs optimal passive beamforming. In this scheme, the number of pilot symbols required  is $\tau = M_\mathrm{R} +1$ for each block. 
\end{enumerate}

\begin{figure}
\vspace*{-1.1em}
\centering
\includegraphics[width=0.29\textwidth]{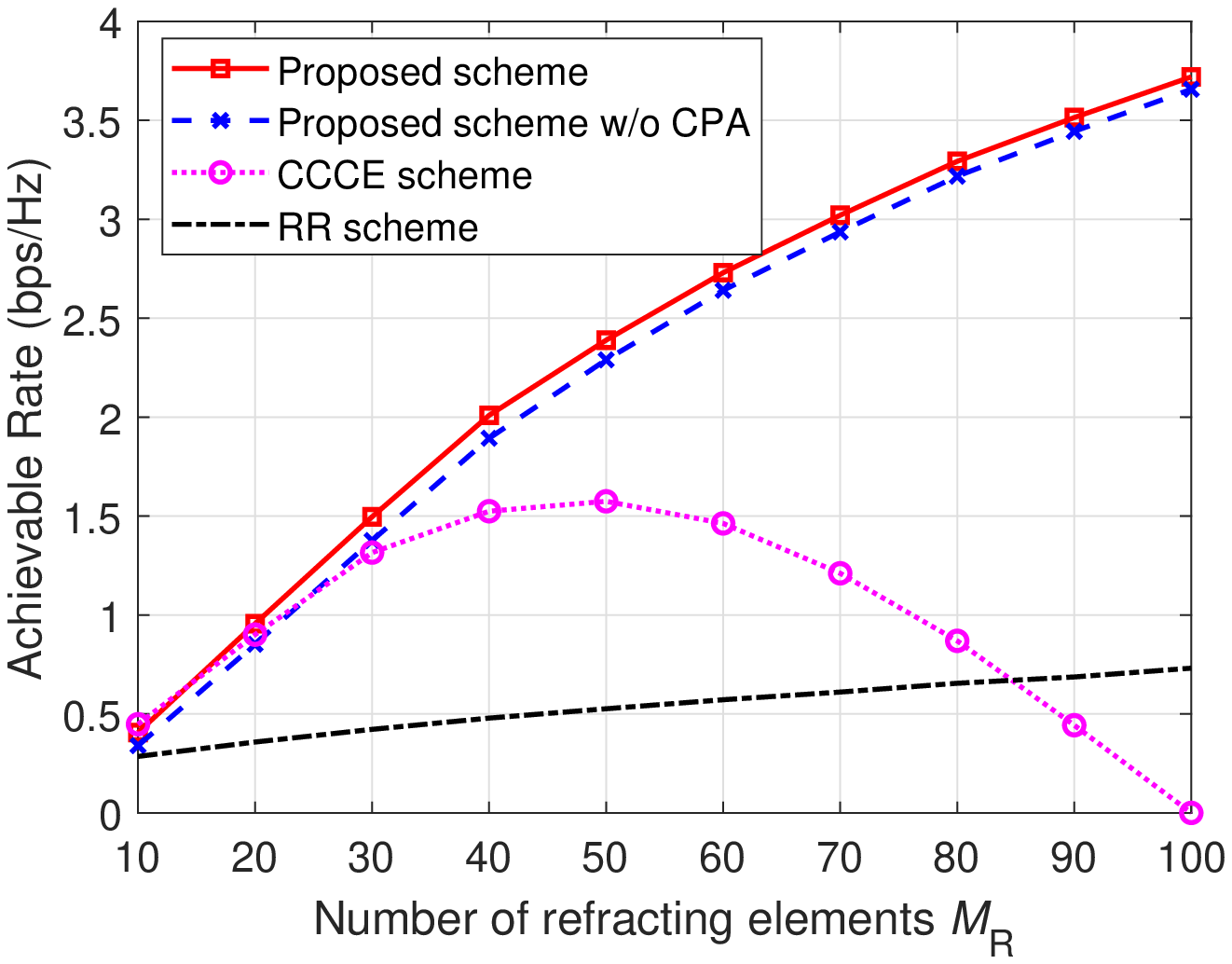}
\vspace*{-0.75em}
\caption{Achievable rate versus $M_\mathrm{R}$.}
\vspace*{-2.1em}
\label{mr}
\end{figure}

In Fig. \ref{mr}, we show the achievable rate versus the number of refracting elements $M_\mathrm{R}$, with $P_t = 31$ dBm and $N_1 = 10$. It is observed that the achievable rates of the proposed scheme with/without CPA and the RR scheme increase with $M_\mathrm{R}$, while the proposed scheme has a much higher increasing rate. This is because the proposed scheme estimates the cascaded channel more efficiently over time and thus achieves higher passive beamforming gains. In contrast, the achievable rate of the CCCE scheme first increases and then substantially decreases with $M_\mathrm{R}$, due to the increasing training overhead with $M_\mathrm{R}$ for the cascaded channel estimation.

\begin{figure}
\centering
\includegraphics[width=0.29\textwidth]{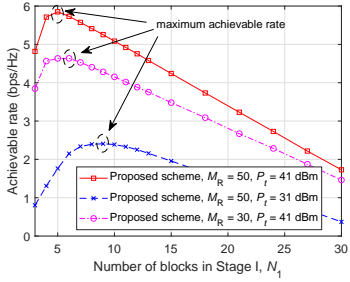}
\vspace*{-0.5em}
\caption{Achievable rate versus $N_1$ for the proposed scheme.}
\vspace*{-2em}
\label{np}
\end{figure}

In Fig. \ref{np}, we show the achievable rate of the proposed scheme versus the number of blocks assigned to Stage~I, $N_1$, where the total number of blocks is fixed to $N=30$. It is observed that there exists a trade-off in the time/block  allocation between the two stages for maximizing the achievable rate.
Moreover, the optimal $N_1$ for rate maximization is shown to decrease with the increasing of  $M_\mathrm{R}$ and/or $P_t$ in general. This is expected since the average power of the IRS-refracted link increases with both $M_\mathrm{R}$ and $P_t$, thus requiring smaller $N_1$ to achieve sufficiently accurate estimation for $\{f_d, \psi_\mathrm{x},\angle \beta \}$. As a result, the proposed scheme can assign more time/blocks to Stage~II so as to enjoy the high passive beamforming gain for data transmission, which thus achieves improved throughput of the system.

\section{Conclusion}
In this paper, we propose a new transmission protocol for efficient channel estimation and refraction design in the IRS-assisted communication with high-mobility user.
The cascaded BS-IRS-user channel is first estimated with the Doppler effect compensated.
Then, the BS-user fast fading  channel is estimated and  the IRS refraction is adjusted  over time accordingly to align it with the cascaded channel, thus achieving high passive beamforming gain as well as converting the overall  channel from fast to slow fading.
Simulation results demonstrate the superior performance of our proposed scheme against benchmark schemes.

\appendix
\section{}
At the end of Stage~I, the user collects the estimates of $\{\bar{h}^{\left( n \right)}_r\}^{N_1}_{n=1}$ over $N_1$ blocks, which can be expressed in a vector form as 
\begin{align}\label{est_s1}
\hat{\bar{\mathbf{h}}}_r 
=& \bar{\mathbf{h}}_r+ \bm{\epsilon} \nonumber\\
=& \beta \bm{\Gamma}\left(f_d\right) \bar{\mathbf{V}} \boldsymbol{\mathrm{s}}^* \left(\psi_\mathrm{x}\right) + \bm{\epsilon},
\end{align}
where $\bar{\mathbf{h}}_r = \left[\bar{h}^{\left( 1 \right)}_r , \ldots, \bar{h}^{\left( N_1 \right)}_r\right]^T \in \mathbb{C}^{N_1 \times 1}$, $\bm{\Gamma}\left(f_d\right) = \operatorname{diag}\left(e^{j 2 \pi f_d T_c}, \ldots, e^{j 2 \pi f_d N_1 T_c}\right) \in \mathbb{C}^{N_1 \times N_1}$ denotes the phase rotation matrix due to the Doppler frequency, $\bar{\mathbf{V}} =  \left[\bar{\mathbf{v}}^{\left(1 \right)}, \ldots, \bar{\mathbf{v}}^{\left( N_1 \right)}\right]^{T} \in \mathbb{C}^{N_1 \times M_\mathrm{R}}$ denotes the initial training refraction matrix, and $\bm{\epsilon} \in \mathbb{C}^{N_1 \times 1}$ denotes the estimation error of $\bar{\mathbf{h}}_r$, whose $n$-th entry with $n\in \mathcal{N}_\mathrm{I}$ is given by the second entry of $\mathbf{\Theta}^\dagger \mathbf{z}^{\left(n\right)}$ in (\ref{est_bl}). 
Based on (\ref{est_s1}), we can jointly estimate $\{ \beta, f_d, \psi_\mathrm{x} \}$ by the maximum likelihood (ML) estimation.  
Specifically, the negative log-likelihood function (after omitting irrelevant 
terms) of $\hat{\bar{\mathbf{h}}}_r$ with given $\{ \beta, f_d, \psi_\mathrm{x} \}$ can be obtained as 
\begin{equation}\label{mle_met}
\Lambda\left( \hat{\bar{\mathbf{h}}}_r \mid \beta, f_d, \psi_\mathrm{x} \right) = \left\|\hat{\bar{\mathbf{h}}}_r -\beta \bm{\Gamma}\left(f_d\right) \bar{\mathbf{V}} \boldsymbol{\mathrm{s}}^* \left(\psi_\mathrm{x}\right)\right\|.
\end{equation}
For notation convenience, we define $\bm{\xi} \in \mathbb{C}^{N_1 \times 1} = \bm{\Gamma}\left(f_d\right) \bar{\mathbf{V}} \boldsymbol{\mathrm{s}}^* \left(\psi_\mathrm{x}\right)$.
Note that for given $f_d$ and $\psi_\mathrm{x}$, the minimizer of (\ref{mle_met}) with respect to $\beta$ is given by
\begin{equation}\label{mle_2}
\hat{\beta} = 
\left( \bm{\xi}^H   \bm{\xi} \right)^{-1}
\bm{\xi}^H 
\hat{\bar{\mathbf{h}}}_r.
\end{equation}
Substituting (\ref{mle_2}) into (\ref{mle_met}), the ML estimates of $ \{f_d, \psi_\mathrm{x} \}$ are obtained as
\begin{equation}\label{mle_1}
\{\hat{f}_d, \hat{\psi}_\mathrm{x} \} = \arg \max _{f_d, \psi_\mathrm{x}} \left(\hat{\bar{\mathbf{h}}}_r\right)^H 
\bm{\xi}  
\left( \bm{\xi}^H   \bm{\xi} \right)^{-1}
\bm{\xi}^H 
\hat{\bar{\mathbf{h}}}_r.
\end{equation}
A two-dimensional grid-based search can be performed to obtain $\{\hat{f}_d, \hat{\psi}_\mathrm{x} \}$ according to (\ref{mle_1}).
Then the estimated channel gain $\hat{\beta}$ is obtained by substituting $\{\hat{f}_d, \hat{\psi}_\mathrm{x} \}$ into (\ref{mle_2}).



\end{document}